# A comparison of graphene, superconductors and metals as conductors for metamaterials and plasmonics

**Philippe Tassin**[1*], **Thomas Koschny**[1], **Maria Kafesaki**[2] **and Costas M. Soukoulis**[1,2]

[1] Ames Laboratory—U.S. DOE and Department of Physics and Astronomy, Iowa State University, Ames, IA 50011, USA

[2] Institute of Electronic Structure and Lasers (IESL), FORTH, 71110 Heraklion, Crete, Greece

[*] e-mail: tassin@ameslab.gov

**Recent advancements in metamaterials and plasmonics have promised a number of exciting applications, in particular at terahertz and optical frequencies. Unfortunately, the noble metals used in these photonic structures are not particularly good conductors at high frequencies, resulting in significant dissipative loss. Here, we address the question of what is a good conductor for metamaterials and plasmonics. For resonant metamaterials, we develop a figure-of-merit for conductors that allows for a straightforward classification of conducting materials according to the resulting dissipative loss in the metamaterial. Application of our method predicts that graphene and high-$T_c$ superconductors are not viable alternatives for metals in metamaterials. We also provide an overview of a number of transition metals, alkali metals and transparent conducting oxides. For plasmonic systems, we predict that graphene and high-Tc superconductors cannot outperform gold as a platform for surface plasmon polaritons, because graphene has a smaller propagation length-to-wavelength ratio.**

Metamaterials and plasmonics, two branches of the study of light in electromagnetic structures, have emerged as promising scientific fields. Metamaterials are engineered materials that consist of subwavelength electric circuits replacing atoms as the basic unit of interaction with electromagnetic radiation[1-3]. They can provide optical properties that go beyond those of natural materials, such as magnetism at terahertz and optical frequencies[4-6], negative index of refraction[7-9], or giant chirality[10]. Plasmonics exploits the mass inertia of electrons to create propagating charge density waves at the surface of metals[11-12], which may be useful for intrachip signal transmission, biophotonic sensing applications, and solar cells, amongst others[13-15].

Unfortunately, although metamaterials and plasmonic systems promise the harnessing of light in unprecedented ways, they are also plagued by dissipative losses—probably the most important challenge to their applicability in real-world devices. In metamaterials, this results in absorption coefficients of tens of decibels per wavelength in the optical domain[16]. In plasmonic systems, dissipative loss is reflected in the limited propagation length of surface plasmon polaritons (SPPs) on the surface of noble metals[17-18]. These losses originate in the large electric currents, leading to significant dissipation in the form of Joule heating, and enhanced electromagnetic fields close to the metallic constituents, leading to relaxation losses in the dielectric substrates on which the metallic elements are deposited. It must be borne in mind that even if the loss tangent of the constituent materials is small, significant losses still occur because the loss channels are driven by large resonant fields. Focusing on terahertz frequencies and higher, loss is dominated by dissipation in the conducting elements, even if noble metals with relatively good electrical properties (e.g., silver or gold) are used.

It has been proposed to reduce the loss problem by replacing noble metals by other material systems[19], e.g., graphene[20-21] or high-temperature superconductors[22]. Both material systems are known to be good conductors, at least for direct currents, and merit further investigation for use in metamaterials or plasmonic systems.

In this work, we answer the question of what is a good conductor for use in metamaterials and in plasmonics. Should it have small or large conductance? Does the imaginary part of the conductivity (or real part of the permittivity, for that matter) improve or worsen the loss? Different applications, e.g., long-range surface plasmons or metamaterials with negative permeability, require conductors with different properties. For resonant metamaterials, we derive a figure-of-merit measuring the dissipative loss that contains the conducting material's properties—the resistivity— and certain geometric aspects of the conducting element. We apply this figure-of-merit to compare graphene, high-$T_c$ superconductors, transparent conducting oxides, transition and alkali metals, and some metal alloys. For plasmonics systems, we use the propagation length to surface plasmon wavelength ratio as the measure of loss performance, and we evaluate graphene as a platform for surface plasmons.

**A figure-of-merit for conductors in resonant metamaterials**

The metamaterials we consider here consist of an array of subwavelength conducting elements; it is for this type of structure that an effective permittivity and permeability makes sense[23-26]. This allows modelling each individual element as a quasistatic electrical circuit described by an RLC circuit. This is not the most general case, as some reported phenomena in metamaterials require more intricate circuits[27-30], but it was proven that it can well capture the physics of the most popular elements, such as split rings and wire pairs[31].

Our analysis starts with describing the electrical current flowing in the metallic circuit of each meta-atom. Subsequently, we calculate the permeability of the metamaterial and the dissipated power by summing the Joule heat loss for each circuit[32] (see the methods section for a detailed derivation). Expressed in dimensionless quantities, we find that the dissipated power as a fraction of the incident power can be cast in the following form:

$$\Pi = \frac{\text{dissipated power per unit cell}}{\text{incident power per unit cell}} = 2\pi \left(\frac{a_k}{\lambda_0}\right) \frac{F\tilde{\omega}^4 \zeta}{\left[\tilde{\omega}^2(1+\xi)-1\right]^2 + \left(\tilde{\omega}\zeta + \tau\tilde{\omega}^5\right)^2}, \quad (1)$$

In equation (1), $\tilde{\omega} = \omega/\omega_0$ is the renormalized frequency, where $\omega_0 = (LC)^{-1/2}$ is the resonance frequency of the quasistatic circuit, $F$ is the filling factor of the metal in the unit cell, $\zeta = \text{Re}(R)/\sqrt{L/C}$ is the "dissipation factor," $\xi = -\text{Im}(R)/(\tilde{\omega}\sqrt{L/C})$ is the "kinetic inductance factor," and $\tau$ is a parameter describing radiation loss. $a_k$ is the metamaterial's unit cell size along the propagation direction and $\lambda_0$ is the free-space wavelength. We will discuss the physical significance of these parameters in the following.

It is interesting to note that the dissipated power fraction quantifying the dissipative loss depends on just four independent, dimensionless parameters:

1. the filling factor, $F$;
2. the radiation loss parameter, $\tau$;
3. the dissipation factor, $\zeta$ (proportional to the real part of the resistivity); and
4. the kinetic inductance factor, $\xi$ (proportional to the imaginary part of the resistivity).

The filling factor and the radiation loss parameter depend only on purely geometric variables, such as the area of the circuit and the geometric inductance, but not on the material properties of the conductor. Thus, for a certain geometry (say, split rings or fishnet), $F$ and $\tau$ are fixed. This means we can limit this study to how the dissipated power depends on $\zeta$ and $\xi$, the only two parameters that depend on the specific conducting material used. In figure 1a, we have plotted a contour plot of the dissipated power fraction as a function of these two parameters of interest. As our aim is to design metamaterials with negative permeability ($\mu$), we calculated the dissipated power at the frequency where $\mu(\omega) = -1$. Apart from the uninteresting regions with very high dissipation factor and/or kinetic inductance factor (top and rightmost regions), we see that the contours of equal dissipated power are almost vertical, i.e., the dissipative loss depends, to a good approximation, only on the dissipation factor. Therefore, we can replot the dissipated power fraction as a function of the dissipation factor (figure 1c); the different blue curves in this figure represent the dissipated power fraction for several values of the achieved permeability. We observe that smaller $\zeta$—i.e., smaller real part of the resistivity—leads to lower power dissipation, even though smaller resistivity implies quasistatic circuits with sharper resonances.

This behaviour can be understood from examining figure 1b. For large dissipation factors (metamaterials made from high-resistivity materials), the resonance is highly damped; the peak loss at the resonance frequency is relatively low and the resonance is too shallow to allow for permeability $\mu = -1$. With decreasing dissipation factor, the resonance becomes sharper; the dissipated power at the resonance frequency increases (red curve in figure 1c), but for a given design goal of the permeability (e.g., $\mu = -1$), the resonance may be probed farther from the resonance peak, effectively leading to smaller dissipated power. When the dissipation factor is further decreased, the linewidth becomes limited by the radiation loss, which implies that the current in the circuit does not further increase. From this point on, the resonance does not become any stronger when the dissipation factor is further decreased (i.e., when a better conductor is used). Therefore, one cannot achieve arbitrarily low permeability, but, on the contrary, there is a geometrical limit on the strongest negative permeability that can be achieved with a certain structure. Since the induced current in the circuit now becomes constant when further decreasing the dissipation factor, the dissipated power at the working frequency continues to decrease linearly with the dissipation factor, because the Joule heating is proportional to the resistivity.

In a similar way, we can show that the kinetic inductance factor determines the frequency saturation due to the kinetic inductance[33, 31]. The reader is referred to the Supplementary Material.

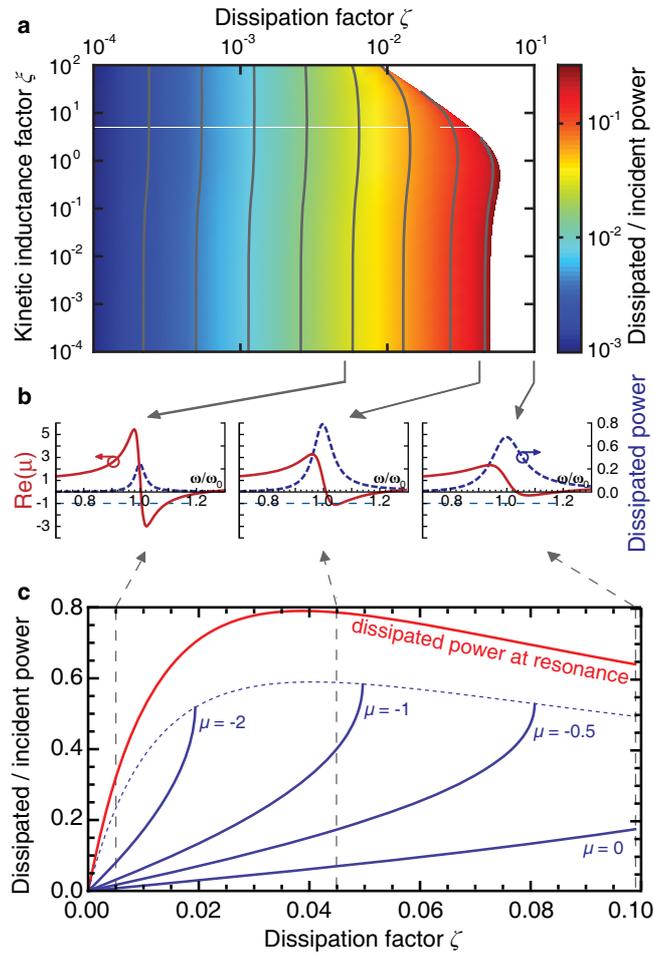

**Figure 1 | Dissipated power in a metamaterial with $F$ = 0.37 and $\tau$ = 0.039; quantities are calculated for the slab-wire pair of figure 2a. a,** Contour plot of the dissipated power (calculated at the operating frequency where $\mu(\omega) = -1$) as a function of the dissipation factor $\zeta$ and the kinetic inductance factor $\xi$. Apart for the uninteresting high dissipation/high kinetic inductance region, the contours are vertical, indicating that the dissipated power depends, to a good approximation, only on the dissipation factor. **b,** Resonance shapes of the magnetic permeability (red lines) and the dissipated power (blue dashed lines) for different dissipation factors. **c,** Dissipated power as a function of the dissipation factor. The red line indicates the peak dissipative loss at the resonance frequency. The blue lines represent the dissipative loss at constant permeability. The dashed blue line indicates the cutoff; for higher dissipation factor the desired permeability can no longer be achieved.

In summary, the dissipative loss in resonant metamaterials can be determined from a single dimensionless parameter—the dissipation factor. From figure 1c, it can be observed that the dissipated power is a monotonic function of the dissipation factor, even approaching a linear function for small $\zeta$. This unambiguously establishes the dissipation factor $\zeta$ as a good figure-of-merit for conducting materials in resonant metamaterials. Whenever a new conducting material is proposed, the dissipation factor allows for a quick and straightforward assessment of the merits of this conducting material for use in the current-carrying elements of resonant metamaterials.

We conclude that resonant metamaterials benefit from conducting materials with smaller real part of the resistivity. However, when comparing conducting materials of which samples of comparables thickness cannot be fabricated, the geometrical details in the dissipation factor become important. This will be essential when we investigate the two-dimensional conductor graphene in the next section. Note that we have derived the loss factor for negative-permeability metamaterials, but it is also applicable for other metamaterials that rely on the resonant response of other polarizabilities, e.g., with negative permittivity and giant chirality.

**Graphene at optical frequencies**

Graphene is a two-dimensional system in which electric current is carried by massless quasiparticles[34-35]. We have seen above that for low-loss resonant metamaterials, we need conducting materials with small real part of the resistivity (to allow for large currents) and small imaginary part of the resistivity (to avoid saturation of the resonance frequency). Band structure calculations and recent experiments indicate that minimal resistivity in the mid-infrared and visible band is achieved for charge-neutral graphene, where the surface conductivity equals the universal value $\sigma_0 = \pi e^2/2h$[36-38].

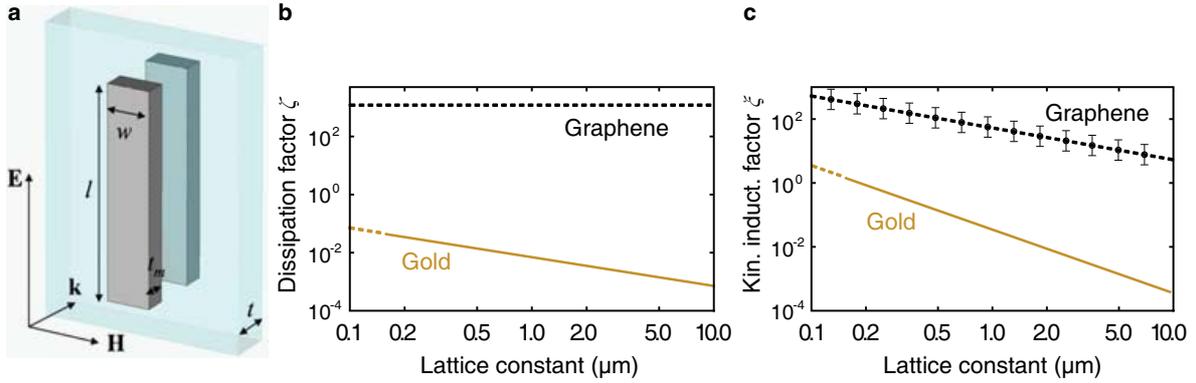

**Figure 2 | Comparison between the loss factors and the kinetic inductance factors of charge-neutral graphene and gold. a,** The slab-wire pair used as an example of a magnetic metamaterial (parameters are provided in the methods section). **b,** The dissipation factor for the slab-wire pair made from graphene and from gold. **c,** The kinetic inductance factor for the slab-wire pair made from graphene and from gold. [The full lines indicate the structure can provide negative permeability ($\mu = -1$); the dashed lines indicate that the structure is beyond the cutoff and that the resonance is too shallow to obtain $\mu = -1$.]

For the slab-wire pair of figure 2a, we have calculated the dissipation factor (figure 2b) and the kinetic inductance factor (figure 2c) for gold and graphene. Gold slab-wire pairs can provide negative permeability, $\mu = -1$ if the lattice constant is larger than 0.15 μm (full line); for smaller lattice constants (dashed line), the dissipation factor increases above the cutoff at which negative permeability cannot be achieved. The dissipation and kinetic inductance factors for graphene are several orders of magnitude larger than those for gold. The dissipation factor of graphene equals 1,200, which is deep into the cut-off region of figure 1a, where the losses are tremendous and the magnetic resonance is highly damped. In addition, the dissipation factor of graphene is scale-invariant; graphene cannot be made a better conductor by making the slab-wire pair larger. We must conclude that graphene is not conducting well enough for use in resonant metamaterials at infrared and visible frequencies.

This observation might not be so surprising given that recent results have demonstrated the optical transmittance through a free-standing graphene sheet to be more than 97%; i.e., graphene has a fairly small interaction cross-section with optical radiation[38]. Many works have ascribed a high bulk conductivity to graphene—obtained by dividing its surface conductivity by the "thickness" of the monatomic layer. This is true, but irrelevant for metamaterial purposes, where it is the total transported current that is important. Graphene might still be useful for metamaterials when it is combined with a metallic structure[39].

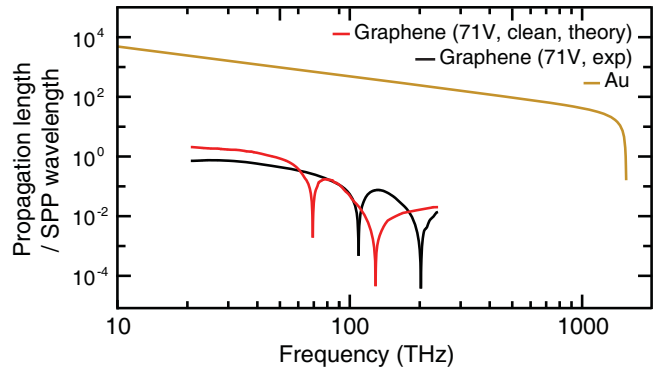

**Figure 3 | Comparison of the plasmonic properties of graphene and gold.** The results for gold are for a 30-nm-thick film at room temperature. The results for graphene are for strongly biased graphene calculated from experimental conductivity data (from ref. 36) and calculated from theoretical data that incorporates electron-electron interactions (from ref. 42). Calculations based on the theoretical data for the conductivity of graphene serve as a best-case scenario, since electron-electron interactions are an intrinsic effect.

There has been recent interest in using graphene as a platform for surface plasmon polaritons (SPP)[40-41,20-21]. For plasmonics, it is desirable to work with biased graphene, because it has much larger kinetic inductance [Im($\sigma$)] than charge-neutral graphene. From Supplementary Figure 4a, we see that biased graphene indeed supports SPPs with wavelengths much smaller than the free-space wavelength. At 30THz, for example, the wavelength of SPPs is 0.2μm. Graphene may thus allow for the manipulation of surface plasmons on a micrometer scale at infrared frequencies. In addition, these SPPs are excellently confined to the graphene surface with submicrometre lateral decay lengths (see Supplementary Figure 4b).

To minimize the loss, we can work in the frequency window just below the threshold of interband transitions where the Drude response of the free electrons is small and the interband transitions are forbidden due to Pauli blocking. The effect of dissipation on SPPs can be best measured by the ratio of their propagation length and wavelength. In figure 3, we show this ratio for gold (yellow curve) and graphene (black curve), calculated from experimental conductivity data obtained by Li *et al*.[36] The propagation length is at best of the order of one SPP wavelength for strongly biased graphene in the infrared. One might object that cleaner graphene samples with smaller Re($\sigma$) might be fabricated in the future. Therefore, as a best-case scenario for SPPs on graphene, we also determined the SPP propagation length based on theoretical data for clean graphene taking into account electron-electron interactions[42-43], which fundamentally limit the conductivity of graphene. We find slightly improved propagation lengths (red curve), but not larger than three SPP wavelengths. Such short propagation lengths will probably be detrimental to most plasmonic applications.

**High-temperature superconductors at terahertz frequencies**

A successful approach towards low-loss microwave metamaterials is the use of type-I superconductors[44]. The microwave resistivity (5 GHz) of niobium sputtered films, for example, is $1.6 \times 10^{-13}$ Ω m at 5 K, roughly 5 orders of magnitude smaller than silver. Unfortunately, this approach is rendered ineffective at terahertz frequencies, because terahertz photons have sufficient energy to break up the Cooper pairs that underlie the superconducting current transport. It has therefore been suggested to use high-temperature superconductors with a larger bandgap.

We know from the above analysis that we must compare the resistivity, which is the geometry-independent part of the dissipation factor (we can leave out the geometrical terms here because metallic and superconducting films of the same thickness can be fabricated). In figure 4, we present a comparison between silver (data from ref. 45) and YBCO (data from ref. 46). We observe that from 0.5 THz to 2.5 THz, both the real part of the resistivity (dissipation) and the imaginary part (kinetic inductance) of YBCO are significantly larger than those of silver. Therefore, we conclude that high-$T_c$ superconductors do not perform better than silver as conducting materials at terahertz frequencies. The reason behind the high resistivity values for YBCO is the specific current transport process occurring in a superconductor. For direct current (DC), the electrons in the normal state are completely screened by the superfluid; hence its zero DC resistivity. At nonzero frequencies, however, the screening is incomplete because of the finite mass of the Cooper pairs. Therefore, the lossy electrons in the normal state contribute to the conductance. In addition, in type-II superconductors, the superfluid has loss mechanisms of its own, like flux creep. Both effects lead to a nonzero resistivity, even at frequencies well below the superconductor's bandgap[44].

For the sake of completeness, we mention that the plasma frequencies of superconductors are not much larger than those of gold or, in other words, that they have similar kinetic inductance. So, at frequencies below the superconductor's bandgap, the dispersion relation of surface plasmons is very close to the light line and superconductors do not support well-confined SPPs.

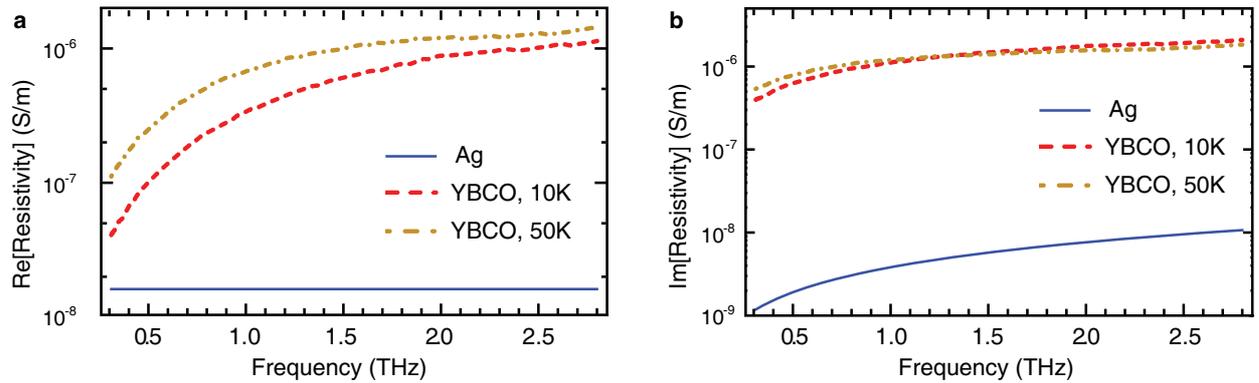

**Figure 4 | Comparison of the superconductor YBCO at 10 and 50K (below the critical temperature of 80K) with silver at room temperature. a,** Real part of the resistivity as a measure of the dissipative loss. **b,** Imaginary part of the resistivity as a measure of kinetic inductance.

**Comparative study of metals and conducting oxides**

In figure 5, we have classified a variety of conducting materials according to their plasma frequency and collision frequency. The collision frequency takes into account all scattering from the conducting electronic states (electron-phonon scattering, interband transitions, etc.) and, therefore, depends on frequency. For most materials, the conductivity in the microwave band (blue symbols in figure 5) is dominated by electron-phonon scattering, although interband transitions may already contribute significantly. At higher frequencies, in the infrared band (red symbols) and the visible band (green symbols), the interband transition scattering becomes larger, in particular close to frequencies matching a transition with high density of states.

At microwave frequencies, silver (■) and copper (◆) have the smallest resistivity; copper is frequently used for its excellent compatibility with microwave technology. Transition metals such as gold (●), aluminium (▲), chromium (⬢), and iridium (⬟) still perform well. The dissipation factors obtained at microwave frequencies are very small and losses in the metals are typically modest (in fact the main loss channel is relaxation losses in the dielectric substrates). In the infrared, the resistivity of copper (◆) is increased tenfold due to interband transitions at 560 nm. The dissipation factors at infrared frequencies are much higher not only due to higher resistivity, but also due to the geometrical scaling of the dissipation factor as shown in figure 2b. Gold (●) performs better than copper in the infrared and is easy to handle experimentally. The reader might notice we have two data points in figure 5 for gold at 1.55 μm; we believe this disparity originates from different grain sizes emphasizing the importance of sample preparation. The best conducting material with the lowest resistivity now becomes silver (■), due to its lowest interband transitions being in the ultraviolet (308 nm). We found ZrN (✖) performs similarly to gold. When further scaling down metamaterials for operation in the visible, the resistivity of most of the abovementioned metals becomes prohibitively high[47] and dissipative losses become too high to obtain, for example, negative permeability. The only reasonably performing metal in the visible is silver (■).

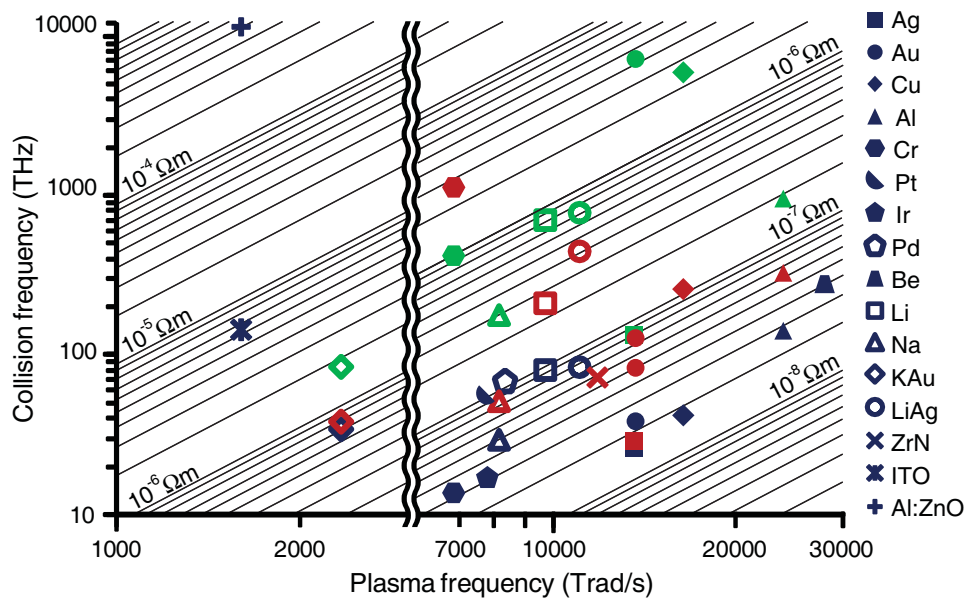

**Figure 5 | Overview of conducting materials classified according to their plasma frequency and collision frequency.** The different symbols indicate different materials (see legend). The collision frequency takes into account all scattering from the conducting electronic states (electron-phonon scattering, interband transitions, etc.) and, therefore, depends upon frequency; blue symbols show material properties at microwave frequencies, red symbols show material properties in the infrared (1.55 μm), and green symbols show material properties in the visible (500 nm). The oblique lines are lines of constant real part of the resistivity and therefore of equal loss performance in metamaterials according to our analysis.

Finding new materials with smaller optical resistivity could have an important impact on the field of metamaterials. We therefore analysed a number of recently proposed alternative conducting materials, e.g., transparent conducting oxides such as indium tin oxide (✻) and Al:ZnO (✚). We find they have a microwave resistivity already two or more orders of magnitude larger than the optical resistivity of silver. Thus, we can rule out these materials; just as for graphene (analyzed above), they interact too weakly with light. Alkali metals suffer less from interband transitions (compare, e.g., lithium (□→□→□) and sodium (△→△→△) with copper (◆→◆→◆)). Unfortunately, their intraband collision frequency is significantly larger and they tend to have a smaller plasma frequency, increasing the average energy lost in each collision. There has also been recent interest in alkali-noble intermetallics with the motivation of combining the low intraband resistivity of the noble metals with the reduced interband transition contribution of the alkali metals[48]. Two characteristic examples are KAu and LiAg. KAu (◇) has its interband transitions far in the ultraviolet and its resistivity increases only slightly from the microwave through the visible; however, its small plasma frequency leads to a relatively large resistivity. On the other hand, LiAg (○) has a larger plasma frequency, but performs badly at higher frequencies because of significant interband scattering. These examples show, nevertheless, the possibility of band engineering to tune the resistivity of alloys[49]. We believe it is worthwhile to continue the research effort to develop better conducting materials, because of the considerable improvement such materials would bring.

**Methods**

The comparative study of conducting materials for resonant metamaterials presented in this work is based on the fact that the dissipative loss in normalized units can be written as a function of two material-dependent parameters—the dissipation factor and the kinetic inductance factor—as expressed in equation (1). This equation is obtained from a quasistatic analysis assuming the conductive elements of the metamaterial to be smaller than the free-space wavelength of the incident radiation. Special attention was paid to the radiation resistance, since its neglect would lead to a circuit model where the dissipated power could become larger than the incident power. The radiation resistance term is obtained from a near-field expansion of the magnetic fields generated by the circuit current, which is again justified by the subwavelength dimensions of the circuit. The details of the derivation of equation (1) are given in the Supplementary Methods and Supplementary Figure 1.

Throughout the manuscript, we exemplified the classification procedure for conducting materials using a particular metamaterial constituent—the slab-wire pair. Nevertheless, the same procedure is applicable for any other metamaterial consisting of subwavelength conducting elements. The slab-wire pair is shown in figure 2a and its dimensions are $l = 2.19\ a_k$, $w = 0.47\ a_k$, $t = 0.5\ a_k$, $t_m = 0.25\ a_k$ ($t_m$ is of course not relevant for two-dimensional conductors such as graphene), $a_E = 2.97\ a_k$, and $a_H = 2.19\ a_k$. The relative permittivity of the substrate is $\varepsilon_r = 2.14$.

We used the simple expressions for the parallel-plate capacitor and for the solenoid inductance; those were shown to provide an adequate description for the slab-wire pair[31]:

$$C = \varepsilon_0 \varepsilon_r \frac{wl}{t}, \quad L = \mu_0 \frac{lt}{a_H}, \quad R = \frac{\rho}{t_m} \frac{2l}{w}. \tag{2}$$

The area enclosed by the circuit is

$$A = lt. \tag{3}$$

This is sufficient to calculate the geometry-dependent term of the dissipation and kinetic inductance factors,

$$\varsigma = \frac{\text{Re}(\rho)}{t_m}\sqrt{\frac{\varepsilon_0}{\mu_0}}\sqrt{\varepsilon_r}\frac{2l\sqrt{a_H}}{t\sqrt{w}},$$

$$\xi = \frac{\text{Im}(\rho)}{t_m}\frac{1}{\tilde{\omega}}\sqrt{\frac{\varepsilon_0}{\mu_0}}\sqrt{\varepsilon_r}\frac{2l\sqrt{a_H}}{t\sqrt{w}}. \quad (4)$$

The filling factor $F$ and the radiation loss parameter $\tau$ can also be calculated:

$$F = \mu_0 A^2 N/L = \frac{lt}{a_k a_E} = 0.37,$$

$$\tau = \frac{1}{6\pi}\frac{\sqrt{\mu_0/\varepsilon_0}}{\sqrt{L/C}}\frac{\omega_0^4 A^2}{c^4} = \frac{1}{6\pi}\frac{1}{\varepsilon_r^{3/2}}\frac{a_H^{5/2} t}{l^2 w^{3/2}} = 0.039. \quad (5)$$

The calculation of the dissipation factor and the kinetic inductance factor for a slab-wire pair made of graphene needs special consideration due to the two-dimensional nature of the current transport. The geometry-dependent terms in $\zeta$ and $\xi$ are calculated in the previous paragraph. The resistivity is obtained from experimental data by Li et al.[36]. The real part of the measured surface conductivity of graphene equals to very good approximation $\sigma_0 = \pi e^2/(2h) = 6.08 \times 10^{-5}$ S/m. The imaginary part is more than 10 times smaller, and, as a consequence, there is a significant uncertainty in its measured value. We have, therefore, fitted two Drude functions to the experimental data: (i) the first provides a lower bound to the measured imaginary part of the conductivity and (ii) the other provides an upper bound (see Supplementary Figure 3). Note that these fits are phenomenological and are unrelated to the Drude-like behaviour of the intraband carriers, since the current transport is dominated by interband carriers in the infrared and the visible. The uncertainty in the imaginary part of the conductivity does not affect the value of dissipation factor, since

$$\text{Re}(\rho) = \frac{\text{Re}(\sigma)}{\text{Re}(\sigma)^2 + \text{Im}(\sigma)^2} \approx \frac{1}{\text{Re}(\sigma)}. \quad (6)$$

However, it does lead to uncertainty in the kinetic inductance factor, indicated with the error bars in figure 2c. The fitted Drude functions are finally used in equations (4) to determine the dissipation factor and the kinetic inductance factors, respectively.

The properties of a surface plasmon polariton $\sim\exp[i(\beta z - \omega t)]$ propagating in the $z$-direction on graphene (dispersion relation in Supplementary Figure 4a, lateral confinement length in Supplementary Figure 4b, and propagation length in figure 3) were calculated from the dispersion relation derived in ref. 20,

$$\beta = \frac{\omega}{c}\sqrt{1 - \left(\frac{2}{\eta_0 \sigma_{//}}\right)^2}, \qquad (7)$$

where $\eta_0$ is the characteristic impedance of free space. The SPP wavelength is obtained from $\lambda_{SPP} = 2\pi/|\text{Re}(\beta)|$, the propagation length by $1/|\text{Im}(\beta)|$, and the lateral decay length by $1/\text{Re}[\sqrt{\beta^2 - (\omega/c)^2}]$. For the conductivity of graphene, $\sigma_{//}$, we have used experimental data for strongly biased ($V_{bias} = 71$ V) graphene from ref. 36. In addition, we have used the theoretical model by Peres *et al.* for the conductivity of graphene to calculate the propagation length of a very clean graphene sample[42]. This theoretical data ignores extrinsic scattering like impurities (which could potentially be removed in cleaner samples), but does account for electron-electron interactions (an intrinsic effect that cannot be removed).

The comparative analysis of metals and conductive oxides in figure 5 is based on experimental data from several sources. In Supplementary Table 1, we list the plasma frequency, the collision frequency, and the resistivity of the metals and the conducting oxides contained in figure 5. References for the experimental data points are also provided in Supplementary Table 1.

**Acknowledgments**

Work at Ames Laboratory was partially supported by the U.S. Department of Energy, Office of Basic Energy Science, Division of Materials Sciences and Engineering (Ames Laboratory is operated for the U.S. Department of Energy by Iowa State University under Contract No. DE-AC02-07CH11358) (theoretical studies) and by the U.S. Office of Naval Research, Award No. N00014-10-1-0925 (study of graphene). Work at FORTH was supported by the European Community's FP7 projects NIMNIL, Grant Agreement No. 228637 (graphene), and ENSEMBLE, Grant Agreement No. 213669 (study of oxides). P. T. acknowledges a fellowship from the Belgian American Educational Foundation.

**Additional information**

The authors declare no competing financial interests.